\documentstyle{article}
\newtheorem{Theorem}{Theorem}[section]
\newtheorem{Definition}[Theorem]{Definition}
\newtheorem{Proposition}[Theorem]{Proposition}
\newtheorem{Lemma}[Theorem]{Lemma}
\newtheorem{Corollary}[Theorem]{Corollary}
\newtheorem{Remark}[Theorem]{Remark}

\def\kasten{\hfil\vrule height6pt width5pt depth-1pt\par }

\def\R{{I\!\! R}}
\def\N{{I\!\! N}}
\def\H{{I\!\! H}}

\def\p{\partial}

\def\half{{1 \over 2}}

\def\a{\alpha}
\def\b{\beta}
\def\d{\delta}
\def\e{\epsilon}
\def\g{\gamma}

\def\l{\lambda}
\def\om{\omega}
\def\s{\sigma}

\def\vp{\varphi}

\def\D{\Delta}

\def\L{\Lambda}
\def\Om{\Omega}

\def\BC{{\cal B}}

\def\DC{{\cal D}}
\def\EC{{\cal E}}
\def\FC{{\cal F}}

\def\HC{{\cal H}}

\def\LC{{\cal L}}
\def\MC{{\cal M}}

\def\PC{{\cal P}}

\def\SC{{\cal S}}
\def\TC{{\cal T}}
\def\UC{{\cal U}}
\def\W{{\sf W}}
\def\w{{\sf w}}

\def\C{\hbox{\vrule width 0.6pt height 6pt depth 0pt \hskip -3.5pt}C}

\begin{document}
\par\noindent
{\large \bf Models of Local Relativistic Quantum Fields with}
\par\noindent
{\large \bf Indefinite Metric (in All Dimensions)}
\bigskip
\par\noindent
S. Albeverio${}^{1,2,3}$, H. Gottschalk${}^1$, and J.-L. Wu${}^{1,2,4}$
\medskip
\par\noindent
${}^1$ Fakult\"at und Institut f\"ur Mathematik der 
Ruhr-Universit\"at Bochum, D-44780 Bochum, Germany
\par\noindent
${}^2$ SFB237 Essen-Bochum-D\"usseldorf, Germany
\par\noindent
${}^3$ BiBoS Research Centre, Bielefeld-Bochum, Germany; and CERFIM,
Locarno, Switzerland
\par\noindent
${}^4$Probability Laboratory, Institute of Applied Mathematics, 
Academia Sinica, Beijing 100080, P R China

\bigskip\par
\begin{abstract}{
A condition on a set of truncated Wightman functions  is formulated and 
shown to permit the construction of the Hilbert space structure included
in the Morchio--Strocchi modified Wightman axioms. 
The truncated Wightman functions which are obtained by
analytic continuation of the (truncated) Schwinger functions 
of Euclidean scalar random fields and covariant vector 
(quaternionic) random fields constructed via convoluted 
generalized white noise, are then shown to satisfy this 
condition. As a consequence such random fields provide 
relativistic models for indefinite metric quantum 
field theory, in dimension 4 (vector case), respectively in 
all dimensions (scalar case).}
\end{abstract}

\bigskip\par
{}
\bigskip\par
\section*{Introduction}
Since the appearance of gauge theories, it became natural to consider 
(local) quantum field theory(abbr. QFT) in which not all of the Wightman 
axioms are satisfied. Such a consideration was in particular natural
and also necessary for the study of "charged" fields interacting with gauge
fields, because their description conflicts either with locality or with 
positivity(positive definiteness of the set of Wightman functions 
\cite{Jo,StWi} and \cite{BLT}). The physical reason for this 
is that in such theories one must use observables of 
the charged type which obey a Gauss' law(see e.g. Morchio and
Strocchi\cite{MoS}), instead of using the usual local observables. 
Actually, from the study of fields such as e.g. $\a$-gauge type Higgs 
models which do not satisfy positivity(see e.g. \cite{JaS} and references 
therein), it turned out that it is better in general to keep the locality 
condition and to give up the positivity condition. This leads to the so 
called "modified Wightman axioms" of the indefinite metric QFT
(see \cite{S}). The difference between indefinite metric QFT and standard
(i.e. positive metric) QFT is that the axiom of positivity in the latter
is replaced by the so-called "Hilbert space structure condition (HSSC)" in 
the former which permits the construction of Hilbert spaces associated to 
the given collection of Wightman functions. 

In recent years models of Euclidean random fields
of scalar and vector type have been constructed via convolution from
generalized white noise, see e.g. \cite{AGW1,AGW2,AIK,AW} and references 
therein. Furthermore, by analytic continuation,
one can get Wightman functions from the Schwinger functions of such
Euclidean models. The corresponding Wightman functions satisfy the 
relativistic postulates on invariance, spectral property, locality and
cluster property. The positivity condition does not hold, in general, for 
the Wightman functions, in fact in \cite{AGW2}(see also \cite{AGW1}) 
a counterexample was given 
to show that the reflection positivity does not hold for the associated 
Schwinger functions, if the non Gaussian component in the 
generalized white noise is sufficiently strong. 
Hence it is very interesting to see whether the Wightman
functions of such models satisfy the modified Wightman axioms for indefinite
QFT's.

The aim of this paper is to prove that the Wightman functions associated 
with the above mentioned Euclidean models indeed satisfy the Hilbert space
structure condition. Such Euclidean fields provide thus relativistic models 
for indefinite metric QFT's. The technique required to achieve this is 
based on explicit formulae for the truncated Wightman functions.

The paper is organized as follows. In Section 1,
we introduce majorant Hilbert topologies, a necessary and sufficient 
condition called Hilbert space structure condition for the existence 
of a majorant Hilbert topology, and modified Wightman
axioms. In Section 2, we present a sufficient Hilbert space structure 
condition for truncated Wightman functions which implies the Hilbert 
space structure condition for Wightman functions. In Section 3, we introduce 
both scalar and vector Euclidean random fields as convoluted
generalized white noise. We give explicit formulae for their truncated
Wightman functions(we remark that the formulae for the 
vector models we give in Section 3 are written in a way which is different, 
although equivalent, from the one used
in \cite{AIK}). We derive them by following the procedure for
the scalar models in \cite{AGW2}, which makes it possible to prove the
temperedness of the truncated Wightman functions(which is a point left
open in \cite{AIK}). Section 4 is devoted to the verification of the Hilbert 
space structure condition for the models introduced in Section 3.

\section{Majorant Hilbert topologies and modified Wightman axioms}
In this section, we introduce a majorant Hilbert topology structure 
associated with Wightman functions. For an extensive mathematical 
account of such topologies as well as of indefinite inner product spaces,
we refer to the monograph Bogn\' ar\cite{B}.
Here we follow the presentation of \cite{S} and \cite{MoS}. 

Let $d\in\N$ be a fixed space-time dimension and $q\in\N$
be a fixed
number. For any
$n\in\N$, let us denote by $\SC(\R^{dn}, \C^{q^n})$
the Schwartz space of all rapidly decreasing 
$\C^{q^n}$-valued $C^\infty$-functions
on $\R^{dn}$ with the Schwartz topology. Let 
$\SC'(\R^{dn},\C^{q^n})$ denote its topological dual. 
Let us begin by introducing the following axioms for Wightman functions 
$\{ \W _n\}_{n\in\N_0}$(with $\W_0=1$ for simplicity):

{\bf Axiom I}(Temperedness) For any $n\in\N$, the $n$-point function 
$\W _n(x_1, \cdots, x_n),$ $ x_1, \cdots, x_n\in\R^d,$ is a tempered 
distribution, i.e., $\W_n\in\SC'(\R^{dn},\C^{q^n})$.

{\bf Axiom II}(Poincar\' e invariance) There is a 
representation $\TC $ of the proper,
 ~orthochronous Lorentz 
group $\LC ^\uparrow _+(\R^{d})$ (which can be assumed to be irreducible)
acting on $\R^q$, such that
for any $n\in\N$ and any Poincar\' e
transformation $\{a,\L\}\in{\cal P}^\uparrow_+(\R^{d})$, the $n$-point
function
$\W _n(x_1, \cdots, x_n)$ is invariant under $\{a, \L\}$:
$$\TC (\L)^{\otimes n}\W _n(\L ^{-1}(x_1-a), \cdots, \L^{-1} (x_n-a))
=\W _n(x_1,\cdots,x_n),$$
which should be understood ~component wise as follows
\begin{eqnarray*}
\W^{j_1,\cdots,j_n}_n(x_1,\cdots,x_n) &=&
\sum^q_{l_1,\cdots,l_n=1}\TC(\L)^{j_1}_{l_1}\cdots\TC (\L)^{j_n}_{l_n}
\\
&\times& \W^{l_1,\cdots,l_n}_n(\L^{-1}(x_1-a),\cdots,\L^{-1}(x_n-a)).
\end{eqnarray*}

\bigskip

We remark that by Axiom II, every $\W _n$ is actually a distribution in
the difference variables, i.e. there is a tempered distribution $\w_n\in
\SC'(\R^{d(n-1)},\C^{q^n})$ defined as
\begin{eqnarray*}
\w_n(y_1,\cdots,y_{n-1}) := \W _n(x_1,\cdots,x_n) 
\end{eqnarray*}
where $y_j:=x_j-x_{j+1}, 1\le j \le n-1$. For $r\in \N$ we adopt
the conventions in Vol.II of \cite{RS} for the (component wise) 
definition of  the Fourier transform $\hat{ }$ on 
$\SC (\R^{dn},\C^{r})$ and $\SC '(\R^{dn},\C^{r})$, respectively.

\bigskip

{\bf Axiom III}(Spectral condition) For any $n\in\N$, 
the Fourier transform $\hat{\w}_n(q_1, \cdots, q_{n-1})$ is
supported in the backward cones 
$$\{(q_1,\cdots,q_{n-1})\in\R^{d(n-1)}: 
q^2_j \ge 0, q^0_j<0,1\le j \le n-1\},$$ 
where $q_j=(q^0_j,\vec{q_j})\in\R\times\R^{d-1}$, 
and $q^2_j:=\vert q^0_j\vert ^2 - \vert \vec{q_j}
\vert ^2$ is in Minkowski metric. (A different sign convention on
the Fourier transform in most of the physical literature leads to the
interchange of forward and backward cones.)

{\bf Axiom IV}(Locality) For $n\ge 2$, if $(x_{j+1}-x_j)^2<0$ for some 
$j\in \{1,\cdots, n-1\}$, then
$$
\W _n(x_1,\cdots,x_j,x_{j+1},\cdots, x_n) =
\pm t_{(j,j+1)}\W _n(x_1,\cdots,x_{j+1},x_j,\cdots,x_n).
$$
Here $+$ corresponds to integer spin of $\TC$, whereas $-$ corresponds
to half-integer spin \cite{StWi}. $t_{(j,j+1)}$ acts on $\W_n=(\W_n^{l_1,\ldots,l_n})
_{l_1,\ldots,l_n=1,\ldots,q}$ by
transposing the ~indexes $l_j$ and $l_{j+1}$.
\bigskip

Let $\underline{\cal S}$ be the Borchers algebra over
$\SC(\R^d,\C^{q})$, namely,
$$\underline{\cal S}:=\{F=(f_0,f_1,\cdots): f_0\in\C,f_n\in\SC(\R^{dn},
\C^{q^n}), \, n\in\N\}
$$
with addition and multiplication given as follows
$$F+G=(f_0+g_0,f_1+g_1,\cdots)$$
$$F\otimes G =((F\otimes G)_0,(F\otimes G)_1,\cdots)$$
where $(F\otimes G)_n:=\sum_{j+l=n}f_j\otimes g_l, n=0,1,2,\cdots$. The topology
on $\underline{\cal S}$ is the direct sum topology induced by the
Schwartz topology of $\SC(\R^d,\C^q)$. Setting
$$\underline{\W}(F):= \sum^\infty_{n=0}\W _n(f_n)
$$
where $\W _0(f_0)=1\cdot f_0$ is the product of the complex numbers $1$ 
and $f_0$, then $\underline{\W}$ is a linear functional on 
$\underline{\cal S}$, called Wightman functional. Furthermore, for
$F=(f_0,f_1,\cdots)\in\underline{\cal S}$, we define its involution by
$F^*=(f^*_0,f^*_1,\cdots)$, where
$f^*_n(x_1,\cdots,x_n):=\stackrel{\leftarrow}{r}\overline{f_n(x_n,\cdots,x_1)}$,where 
$\stackrel{\leftarrow}{r}$ 
acts on $f=(f^{l_1,\ldots,l_n})_{l_1,\ldots,l_n=1,\ldots,q}$ by reversing 
the order of the ~indexes and the bar denotes complex conjugation. Then 
$\underline{\W}$ determines a sesquilinear form on $\underline{\cal S}$
as follows
\begin{equation}
\label{1.1eqa}
<F,G>_W:=\underline{\W}(F^*\otimes G), 
\, \, \, \, F,G\in\underline{\cal S}.
\end{equation}
 
Clearly, $<\cdot,\cdot>_W$ is hermitian if the Wightman functions $\W _n,
n\in\N,$ satisfy the \underline{hermiticity condition}
$$\W _n(x_1,\cdots,x_n)=\stackrel{\leftarrow}{r}\overline{\W _n(x_n,\cdots,x_1)}, \, \,
n\in\N.$$
Hereafter we assume this condition for simplicity.

Now set
$${\cal N}_W:=\{F\in\underline{\cal S}: <F,G>_W=0, \, 
\forall G\in\underline{\cal S}\}
$$
which is the kernel of $<\cdot,\cdot>_W$, then the quotient space 
$${\cal D}:= \underline{\cal S}/{\cal N}_W$$
is well defined as an indefinite inner product space(cf. \cite{B}
for this notion)
with respect to the indefinite inner product induced by 
$<\cdot,\cdot>_W$(we denote the induced product by the same notation).
In general, $({\cal D}, <\cdot,\cdot>_W)$ can not be a pre-Hilbert space. 
However, we may specify some Hilbert inner product which dominates
$<\cdot,\cdot>_W$. To this end, we introduce the following notion

\begin{Definition}
\label{1.1def}
By \underline{a majorant Hilbert topology} $\tau$ of 
$<\cdot,\cdot>_W$ on ${\cal D}$ 
we mean a topology determined by a Hilbert inner product $(\cdot,\cdot)$
on ${\cal D}$ such that
\begin{equation}
\label{1.2eqa}
\vert <F,G>_W\vert\le (F,F)^\half(G,G)^\half, \, \,  F,G \in{\cal D}.
\end{equation}
\end{Definition}

\begin{Remark}\label{1.1rem}
An important property of a majorant Hilbert topology $\tau$
is that from (\ref{1.2eqa}), we have 
$$F^{(n)} \stackrel{\tau}{\to} F \Longrightarrow 
<F^{(m)},F^{(n)}>_W \to <F,F>_W.$$  
Namely, the topology $\tau$ is strong enough for $\tau$-convergence
to imply convergence of all the corresponding Wightman functions 
with respect to the inner product $<\cdot,\cdot>_W$.  \kasten
\end{Remark}

From Definition \ref{1.1def}, $({\cal D}, (\cdot,\cdot))$ is a pre-Hilbert 
space.
Setting ${\cal H}:= {\overline{\cal D}}^{(\cdot,\cdot)}$, then $({\cal H},
(\cdot,\cdot))$ is a Hilbert space. By the known Riesz theorem, 
(\ref{1.2eqa}) implies that there exists a bounded self-adjoint operator $T$
on ${\cal H}$, hereafter called \underline{the metric operator} corresponding 
to $(\cdot,\cdot)$, such that 
$$<F,G>_W = (F,TG), \, \, \, F,G \in {\cal H}.$$  
Moreover, such an operator can be chosen to be non-degenerate, i.e., $T$ 
fulfills ${\cal H}_T$ = \{0\}, where ${\cal H}_T:=\{F\in{\cal H}:
(F,TG)=0, \, \, \forall G \in {\cal H}\}$ is a Hilbert subspace of ${\cal H}$.
Actually, suppose that ${\cal H}_T \ne \{0\}$. We remark that $T({\cal
H}_T)=\{0\}$, thus the following Hilbert inner product
$$(F,G)_1 := (F, (1-{\bf P}_T)G)$$
also determines a majorant Hilbert topology $\tau _1$ of $<\cdot,\cdot>_W$ 
on ${\cal D}$, where
${\bf P}_T: {\cal H} \to {\cal H}_T$ is the projection. Clearly, the  
metric operator $T_1:=(1-{\bf P}_T)^{-1}T$ is non-degenerate(corresponding
to $(\cdot,\cdot)_1$). We call such a $\tau _1$ a non-degenerate majorant 
Hilbert topology. 

In addition, such a procedure of removing the degeneracy of
metric operators also removes the ~nontrivial ideals of the Borchers
algebra $\underline{\SC}$ arising from properties like locality and
spectral conditions of Wightman functions. On the other hand, 
we can (well) define a field operator(i.e., an operator
valued distribution) $\phi(f)$ on the dense domain $\DC \subset
{\cal H}$ for any $f \in \SC (\R^d,\C^q)$ as follows
$$(\phi(f))(G) := F_f \otimes G +{\cal N}_W, \, \, G \in {\cal D}$$
where $F_f:=(0,f,0,\cdots) \in \underline{\cal S}$, with the property that
\begin{eqnarray}
\label{XX}
\W _n(x_1,\cdots,x_n) &=& (\Om, T\phi(x_1)\cdots\phi(x_n)\Om)  \nonumber \\
&=& (\phi(x_j)\cdots\phi(x_1)\Om,T\phi(x_{j+1})\cdots\phi(x_n)\Om)  \nonumber \\   
&=& <\phi(x_j)\cdots\phi(x_1)\Om,\phi(x_{j+1})\cdots\phi(x_n)\Om>_W
\end{eqnarray}
where $\Om:=(1, 0,\cdots) + {\cal N}_W$. Clearly, $<\Om,\Om>_W > 0$.

Since $\DC$ by definition of $\phi $ is a multiplication core for the field
$\phi$, products of field operators $\phi(f)\phi(g)$,$f,g\in \SC (\R^d,\C^q)$, are
well-defined on $\DC$. By Axiom IV and equation (\ref{XX}) the field
operators $\phi(f)$ are local in the sense that 
$$\left[ \phi(f),\phi(g)\right] _\mp=0$$
if the support of the test functions $f,g\in \SC (\R^d,\C^q)$ is space-
like separated. Here $[\cdot,\cdot]_\mp$ stands for the commutator if
the spin of $\TC $ is integer and for the anticommutator ~otherwise 
(cf. Axiom IV).

By equation (\ref{XX}) and the hermiticity condition, we conclude that
the field operator $\phi $ is \underline{$T$-symmetric} in
the sense that for $f \in \SC (\R^d,\C^q)$
$$T\phi (\overline{f})^*T^{-1} = \phi(f).$$
Furthermore, from the action of $\PC^\uparrow_+$ (via $\TC$) 
on the test function spaces $\SC(\R^{dn},\C^{q^n})$ we get a representation
$\UC $ of the proper orthochronous Poincar\'e group
$\PC ^\uparrow _+(\R^d)$ by \underline{$T$-unitary} operators defined on
the common dense domain $\DC$, where,
by definition, an operator $\UC(a,\L)$ on $\HC$ is called $T$-unitary, if
$$ T\UC(a,\L)^*T^{-1}=\UC(a,\L)^{-1}.$$
The field $\phi$ transforms under $\UC $ as
\begin{equation}
\label{YY}
 \UC(a,\L)\phi (x) \UC(a,\L)^{-1} = \TC(\L)\phi(\L^{-1}(x-a)).
\end{equation}
Furthermore, the spectral condition in Axiom III is equivalent to
the following condition
\begin{equation}
\label{ZZ}
\int_{\R^d}(F,T\UC(a,1)G)e^{iqa}da = 0 \, \, ,  F,G \in \DC 
\end{equation}
if $q\notin\{q\in\R^d: q^2\ge 0 \, , q^0<0\}$.

Now we define a "Krein topology" for ${\cal D}$ as follows

\begin{Definition}
\label{1.2def}
A non-degenerate majorant Hilbert topology $\tau$ on ${\cal D}$ is
called \underline{a Krein topology} if 
${\cal H} := {\overline{\cal D}}^{\tau}$
is maximal. Namely, if $\tau _1$ is another non-degenerate majorant
Hilbert topology on ${\cal D}$ such that ${\overline{\cal D}}^{\tau _1}
\supset {\cal H}$, then $\tau = \tau _1$.
\end{Definition}

From Definitions \ref{1.1def} and \ref{1.2def}, it is 
clear that a Krein topology is
a minimal topology to provide maximal information from Wightman
functions and to keep the density of ${\cal D}$ in ${\cal H}$. Moreover,
we have the following result from \cite{MoS}:

\begin{Proposition}
\label{1.1prop}
(Morchio and Strocchi). A majorant Hilbert topology $\tau$ is a Krein
topology iff the corresponding metric operator $T$ has a bounded inverse
$T^{-1}$. Furthermore, such a bounded invertible operator can be chosen
with the property that $T^2 = 1$.
\end{Proposition}

Given a majorant Hilbert topology $\tau$ with the non-degenerate metric
operator $T$ on the Hilbert space $(\HC,(\cdot,\cdot))$, one can always 
find a corresponding Krein topology associated with it. To this end, 
by Proposition \ref{1.1prop}, it is sufficient to find a metric operator
with bounded inverse. In fact, remarking that $T$ is self-adjoint and 
bounded, the absolute operator $\vert T \vert$ is well defined. 
Furthermore, 
$$(F,G)_K := (F,\vert T\vert G), \, \, \, F,G\in\HC$$
determines a new Hilbert inner product whose induced Hilbert topology 
$\tau_K$ is weaker than $\tau$ since
$$(F,F)_K = (F,\vert T\vert F) \le \Vert T\Vert(F,F), \, F\in\HC .
$$ 
On the other hand, we have 
$$<F,G>_W=(F,TG)=(F, (sign \, T)G)_K=:(F,T_KG)_K, \, \, F,G\in\HC.$$
Obviously, $T^{-1}_K=T_K=sign\,T$ is bounded.

\bigskip\par
Concerning the existence of a majorant Hilbert topology, we have the 
following crucial condition from \cite{MoS} and \cite{S}:

\begin{Theorem} 
\label{1.1thm}
(Morchio and Strocchi). Given a collection of Wightman functions
$\{\W _n\}_{n\in \N_0}$, a necessary and sufficient condition for the 
existence of a majorant Hilbert topology is that the following holds:
\bigskip\par
{\bf Axiom V} There is a sequence $\{p_n\}_{n\in\N}$, where $\forall n\in\N,
p_n : \SC(\R^{dn},\C^{q^n}) \to [0,\infty)$ is a Hilbert seminorm, such that

\begin{equation}
\label{1.3eqa}
\vert \W_{m+n}(\vp^*\otimes \eta) \vert \le p_m(\vp)p_n(\eta)
\end{equation}3
for all $\vp \in \SC(\R^{dm},\C^{q^m}), \eta \in \SC(\R^{dn},\C^{q^n}), \, \, 
m,n \in \N$.
\end{Theorem} 

Axiom V is called the \underline{Hilbert space structure condition}. It
is a new axiom for Wightman functions replacing the positivity condition
in the standard QFT. The Axioms I--IV together with the Axiom V are called 
modified Wightman axioms. Such axioms, especially the Hilbert space 
structure condition, were presented and lucidly discussed in \cite{S}.

We remark that, in general, when the Wightman functions do not fulfill the 
positivity condition, one can not expect in general a unique Hilbert space
structure for the states of the theory(the Hilbert space structure depending 
on the choice of Hilbert seminorms in Axiom V). This is at variance with non 
indefinite metric QFT, where the positivity condition guarantees 
the uniqueness of the physical Hilbert space. Uniqueness for 
indefinite metric QFT can perhaps be restored in terms of scattering theory, 
see Remark \ref{1.2rem} below.

Lastly, let us also present the \underline{cluster property} for 
Wightman functions.
A sequence of Wightman functions $\{\W_n\}_{n\in\N_0}$ satisfies the
cluster property if for any $m,n\in\N$ and any space-like $a\in \R^d$(i.e.,
$a^2<0$ in Minkowski metric)
\begin{eqnarray*}
&&\W_{m+n}(\vp_1\otimes\cdots\otimes\vp_mT_{\l a}(\vp_{m+1}\otimes \cdots\otimes
\vp_{m+n}))\stackrel{\l\to\infty}{\longrightarrow} \\
&&\hspace{1in}\W_m(\vp_1\otimes\cdots
\otimes\vp_m)\W_n(\vp_{m+1}\cdots\otimes\vp_{m+n})
\end{eqnarray*}
for $\vp_1,\cdots,\vp_{m+n}\in\SC(\R^d,\C^q)$, where $T_{\l a}$ denotes
the representation of the translation by $\l a$ on $\SC(\R^{dn},\C^{q^n})$.

\begin{Remark}
\label{1.2rem} 
We point out that the cluster property of Wightman functions is
not an item of the modified Wightman axioms, since it does not (directly) 
imply the uniqueness of the vacuum and irreducibility of the field 
algebra as it does in the standard QFT. Nevertheless, the cluster
property can still be looked upon as a genuine expression for the physical 
principle "forces decrease with the (spatial) distance" in
indefinite metric QFT. Especially we expect that also in 
indefinite metric QFT there is a crucial connection between
the cluster property and the possibility of an axiomatic scattering 
theory in such quantum field theories.
\end{Remark}

\section{A sufficient Hilbert space structure condition for truncated
Wightman functions}
Given a sequence of Wightman functions $\{ \W _n\}_{n\in\N_0}$,
$\W_0=1$, $\W_n \in \SC'(\R ^{dn},\C^{q^n})$, the corresponding sequence
of \underline{truncated Wightman functions} $\{ \W_n^T\}_{n\in \N}$, $\W_n^T
\in \SC'(\R ^{dn},\C^{q^n})$, is defined
recursively by the equations
\begin{equation}
\label{2.1eqa}
\W_n(\vp_1\otimes\cdots\otimes \vp_n)=\sum _{I\in \PC ^{(n)}} \epsilon_F (I) 
\prod _{\{j_1,\cdots,j_l\}\in I}\W_l^T(\vp_{j_1} \otimes \cdots \otimes
\vp_{j_l}), \ n\geq 1,
\end{equation}
where $\vp_1, \cdots, \vp_n \in \SC (\R^d,\C^q)$ and $\PC ^{(n)}$ stands
for the collection of all partitions $I$ of $\{ 1,\cdots,n\}$ into
disjoint subsets. For each such subset $\{j_1,\cdots, j_l\}\in I$ we
assume that $j_1<\cdots <j_l$. $\epsilon_F(I)$ stands for the fermionic
parity of the partition $I$, i.e. $\epsilon_F(I):=1$ for (bosonic) integer
spin ${\cal T}$ and 
$$\epsilon_F(I):= \prod_{j<l} \mbox{sign} (\pi_I(l)-\pi_I(j))$$
for (fermionic) half-integer spin ${\cal T}$. For $I=\{ \{j_1^1,\ldots,j_{l_1}^1\},
\ldots,\{j_1^k,\ldots,j_{l_k}^k\}\}$ with $j_1^1<\ldots<j_1^k$,  $\pi_I$
is defined as the permutation which maps $(1,\ldots,n)$ to $(j_1^1,\ldots,j_{l_1}^1,
\ldots,j^k_1,\ldots,j^k_{l_k})$. By the nuclear theorem the sequence of
truncated Wightman functions is determined uniquely by the sequence of
Wightman functions and vice versa.

Since the truncated Wightman functions of the models introduced in
Section 3 below are much simpler objects than the Wightman
distributions themselves, it seems natural to ask for a sufficient
condition on the truncated Wightman functions which implies the Hilbert
space structure condition(HSSC) for the Wightman functions, as it was
introduced in Section 1. The aim of this section is to deduce such a
\underline{HSSC for truncated Wightman functions}
which is then verified in Section 4 for both models of Section 3.

Let us first introduce a special system of Schwartz norms 
$\{\|\cdot\|_{K,N}\}_{K,N\in \N_0} $ on the spaces $\SC (\R ^{dn},\C^{q^n})$,
$n\in \N$, by
$$ \| \vp \| _{K,N} := \sup _{\begin{array}{c} {\scriptstyle x_1,\cdots,x_n 
\in \R^d} \\ 
{\scriptstyle 0\leq|\a_1|,\cdots,|\a_n|\leq K }\end{array}} \left| 
\prod _{l=1}^n
(1+|x_l|^2)^{N/2} D^{\a_1\cdots\a_n}\vp (x_1,\cdots,x_n)\right|\ , $$
for $K,N \in \N_0$ and $\vp \in \SC(\R^{dn},\C^{q^n})$. Here the absolute 
$|\cdot |$ is induced by the scalar product $<\cdot,\cdot>_E^{\otimes n}$ on 
$\C^{q^n}\cong (\C^q)^{\otimes n}$ where $<\cdot,\cdot>_E$ stands for the Euclidean 
scalar product on $\C^q$, $\a_1,\cdots, \a_n
\in \N_0^d$ are ~multi--indexes and for $\a_j = (\a_j^0,\cdots,\a_j^{d-1})$ we
have used the notations $|\a_j|=\a_j^0+\cdots+\a_j^{d-1}$ and 
$$D^{\a_1\cdots\a_n}:=D^{\a_1}\otimes\cdots\otimes D^{\a_n}\ ,$$
where
$$ D^{\a_j} := {\p ^{|\a_j|}\over (\p x^0)^{\a_j^0}\cdots (\p
x^{d-1})^{\a_j^{d-1}}} \ .$$

The definition of the Schwartz norms $\|\cdot\|_{K,N}$ clearly implies that
for $m,n\in \N_0$, $\vp \in \SC (\R^{dm},\C^{q^m})$, $\eta \in \SC
(\R^{dn},\C^{q^n})$ we get
\begin{equation}
\label{2.2eqa}
\| \vp \otimes \eta\| _{K,N} = \| \vp \| _{K,N} \ \|
 \eta \| _{K,N} \ .
\end{equation}

The following lemma shows that the Schwartz norms $\|\cdot\|_{K,N}$ are also
well adapted to the operation of taking the tensor product of two
tempered distributions:

\begin{Lemma}
\label{2.1lem}
Let $m,n\in \N$, $K,N\in \N_0$ and $R\in \SC'(\R^{dm},\C^{q^m}), S\in
\SC'(\R^{dn},\C^{q^n})$. If there exist constants $C_R, C_S>0$, such that 
$$|R(\vp)|\leq C_R\| \vp \|_{K,N}, \ \forall \vp \in 
\SC(\R^{dm},\C^{q^m})$$
and
$$|S(\eta)|\leq C_S\| \eta \|_{K,N}, \ \forall \eta \in
 \SC(\R^{dn},\C^{q^n}),$$
then
$$|R\otimes S\ (\chi)|\leq C_RC_S\| \chi \|_{K,N}, \ \forall \chi
 \in \SC(\R^{d(m+n)},\C^{q^{(m+n)}})\ .$$
\end{Lemma}
\noindent {\bf Proof} By Vol.II(see  p.115) of \cite{GV} there exist 
continuous, polynomially bounded functions $F_R:\R^{dm}\to \C^{q^m}$, 
$F_S:\R^{dn}\to
\C^{q^n}$ and polynomials $P_R,P_S$, such that $R=P_R(D)F_R$ and
$S=P_S(D)F_S$ holds in the sense of tempered distributions.
Consequently, for $\chi \in\SC (\R^{d(m+n)},\C^{q^{(m+n)}})$, we get that
$$ R\otimes S \ ( \chi ) = F_R\otimes F_S \ (P_R(-D)\otimes P_S(-D) \ \chi )
\ . $$
The right hand side(RHS) can be rewritten as an integral over $\R^{d(m+n)}$ 
where the integrand is a product of a polynomially bounded function with a 
fast falling function. Thus, the integral converges absolutely and by
Fubini's theorem we get
$$R\otimes S \ ( \chi ) = F_R (P_R(-D)\varrho) = R (\varrho) \ ,$$
where
\begin{eqnarray*}
\varrho (x_1,\cdots,x_m) &:= & S(\chi (x_1,\cdots,x_m,\cdot)) \\
& = & F_S ({\bf 1}_m\otimes P_S(-D)\chi(x_1,\cdots,x_m,\cdot)).
\end{eqnarray*}
Clearly, $\varrho \in \SC(\R^{dm},\C^{q^m})$. Here we denoted the 
identity operation on $\SC(\R^{dm},\C^{q^m})$ by ${\bf 1}_m$. 
Therefore one gets
\begin{eqnarray*}
|R\otimes S\ (\chi )| & \leq & C_R \| \varrho \| _{K,N} \\
&=& C_R \sup _{\begin{array}{c} {\scriptstyle x_1,\cdots,x_n 
\in \R^d} \\
{\scriptstyle 0\leq|\a_1|,\cdots,|\a_n|\leq K }\end{array}}\prod
_{l=1}^m (1+|x_l|^2)^{N/2} \\
&\times & \left| S(D^{\a_1\cdots\a_n}\otimes 
{\bf 1}_n \ \chi  (x_1,\cdots ,x_m,.))\right| \\
&\leq & C_R \sup _{\begin{array}{c} {\scriptstyle x_1,\cdots,x_n 
\in \R^d} \\ 
{\scriptstyle 0\leq|\a_1|,\cdots,|\a_n|\leq K }\end{array}}\prod
_{l=1}^m (1+|x_l|^2)^{N/2} \\
&\times& C_S \sup _{\begin{array}{c} 
{\scriptstyle x_{m+1},\cdots,x_{m+n} 
\in \R^d} \\ 
{\scriptstyle 0\leq|\a_{m+1}|,\cdots,|\a_{m+n}|\leq K }\end{array}}
\prod_{l=m+1}^{m+n} (1+|x_l|^2)^{N/2}\\
&\times & \left|D^{\a_1\cdots\a_m}\otimes D^{\a_{m+1}\cdots\a_{m+n}}
\chi (x_1,\cdots,x_{m+n}) \right| \\
& = & C_RC_S \|\chi \| _{K,N}
\end{eqnarray*}
\kasten
The following theorem gives a sufficient Hilbert space structure
condition on the truncated Wightman functions:

\begin{Theorem}
\label{2.1thm}
 Let $K,N\in \N_0$ and let  $\{a_n\}_{n\in \N}$ be a sequence of 
positive constants, such that for all $n\in \N$ the
truncated $n$-point Wightman function $\W_n^T$ fulfills
\begin{equation}
\label{2.3eqa}
|\W_n^T(\vp)|\leq a_n \| \vp \| _{K,N}, \ \forall \vp \in \SC
(\R^{dn},\C^{q^n}) \ .
\end{equation}
Then the corresponding sequence of Wightman functions fulfills the
Hilbert space structure condition.
\end{Theorem}
We first prove an ~auxiliary lemma:

\begin{Lemma}
\label{2.2lem}
For any sequence of positive constants $\{b_n\}_{n\in \N}$ there exists
a sequence of positive constants $\{c_n\}_{n\in \N}$, such that for all
$m,n\in \N$ the inequality $b_{m+n}\leq c_mc_n$ holds.
\end{Lemma}
\noindent {\bf Proof} Let $c_n:=\max\{\max\{b_j:1\leq j\leq 2n\},1\}$. Then 
for all $m,n\in \N$, we have $b_{m+n}\leq \max\{c_m,c_n\}\leq c_mc_n$.\kasten

\

\noindent {\bf Proof of Theorem \ref{2.1thm}} For $n\in \N$ we define
$$ b_n:=\sum_{I\in\PC ^{(n)}}\prod _{\{j_1,\cdots,j_l\}\in I} a_l \ . $$
For $m,n\in \N$, $\chi \in \SC (\R^{d(m+n)},\C^{q^{(m+n)}})$, we get by inductive
use of (\ref{2.3eqa}) and Lemma \ref{2.1lem}
$$|\W_{m+n}(\chi)|\leq b_{m+n} \|\chi \| _{K,N} \ .$$
Now we take $\chi =\vp^*\otimes\eta$, $\vp \in\SC (\R^{dm},\C^{q^m}),
\eta \in \SC(\R^{dn},\C^{q^n})$, then by (\ref{2.2eqa}) we get
$$|\W_{m+n}(\vp^*\otimes\eta)|\leq b_{m+n}\|\vp\|_{K,N}\|\eta\|_{K,N} \ . $$
On the other hand, by Lemma \ref{2.2lem} there exists a sequence of 
positive numbers $\{c_n\}_{n\in\N}$ such that
\begin{equation}
\label{2.4eqa}
|\W_{m+n}(\vp^*\otimes\eta)|\leq c_m\|\vp\|_{K,N}c_n\|\eta\|_{K,N} \ . 
\end{equation}
By  Vol.IV (see p. 82) of \cite{GV} for $n\in \N$ there is a system $\{
\| \cdot\|'_{K,N}\}_{K,N\in\N_0}$ 
of Hilbert norms on $\SC (\R^{dn},\C^{q^n})$ which
is equivalent to the system of Schwartz norms 
$\{ \| \cdot\|_{K,N}\}_{K,N\in\N_0}$. 
Thus, there is a sequence of positive constants
$\{ d_n\}_{n\in\N}$ such that for the above fixed $K,N\in \N_0$ and suitable 
$K',N'\in \N_0$ 
(depending on $K,N$ and $n$) we get
$$ \|\vp\|_{K,N}\leq d_n\|\vp\|'_{K',N'}, \ \forall \vp \in
\SC (\R^{dn},\C^{q^n})  \ .$$
We now choose Hilbert norms $p_n$ on $\SC(\R^{dn},\C^{q^n})$ as $p_n(\cdot):=
c_nd_n\|\cdot\|'_{K',N'}$. From (\ref{2.4eqa}) we immediately get
the Hilbert space structure condition
$$ |\W_{m+n}(\vp^* \otimes \eta)|\leq p_m(\vp)p_n(\eta) \ . $$
\kasten

Since all truncated Wightman functions $\W_n^T$ are tempered distributions
and are therefore ~continuous with respect to some norm $\|\cdot\|_{K(n),N(n)}$,
it is enough to ~check (\ref{2.3eqa}) for $n$ larger than a
certain number $m\in\N$: We may simply put $K':=\max\{K,K(n):n=1,\cdots,
m\}$, $N':=\max\{N,N(n):n=1,\cdots, m\}$ and by 
$\|\cdot\|_{K',N'} \geq \| \cdot \|_{K,N}$ 
and $\|\cdot\|_{K',N'} \geq \ \|\cdot\| _{K(n),N(n)}, \ n=1,\cdots, m,$ 
we get (\ref{2.3eqa})
for all $n\in \N$ if the numbers $K,N$ are replaced by $K',N'$. In
particular, we get

\begin{Corollary}
\label{2.1cor}
Let $\{\W_n^T\}_{n\in\N}$ be a sequence of truncated Wightman
distributions. If $\ \W_n^T=0$ for all $n$ larger than a certain number
$m\in\N$, then the corresponding sequence of Wightman functions
fulfills the Hilbert space structure condition. \kasten
\end{Corollary}

Since in our models introduced in Section 3, we have 
explicit formulae for
the Fourier transformed truncated Wightman functions rather than for the
truncated Wightman functions themselves, we need the following Fourier
transformed version of Theorem \ref{2.1thm}:
\begin{Corollary}
\label{2.2cor}
Let $K,N$ and $\{ a_n\}_{n\in\N}$ as in Theorem \ref{2.1thm}.
Suppose that $\forall n\in\N$ the Fourier transformed
truncated $n$-point Wightman function $\hat \W _n^T$ fulfills
\begin{equation}
\label{2.5eqa}
|\hat \W_n^T(\vp)|\leq a_n \| \vp \| _{K,N}, \ \forall \vp \in \SC
(\R^{dn},\C^{q^n}) \ .
\end{equation}
Then the sequence of Wightman functions fulfills the
Hilbert space structure condition.
\end{Corollary}
\noindent {\bf Proof} By the basic fact that $\W_n(\vp)=\hat
\W_n(\hat\vp)$ $ \forall \vp \in \SC (\R^{dn},\C^{q^n})$ we only have to
replace the sequence of Hilbert norms $\{ p_n\}_{n\in\N} $ constructed
in the proof of Theorem \ref{2.1thm} by the sequence $\{ \hat
p_n\}_{n\in\N}$ defined as $\hat p_n:=p_n\circ \hat{ }$. Then $\{
\W_n\}_{n\in\N_0}$ fulfills the Hilbert space structure
condition with respect to $\{ \hat p_n\}_{n\in\N}$.\kasten

\section{Relativistic fields from convoluted generalized white noise}
Since the work by Nelson\cite{N}, the problem of constructing Markovian
or reflection positive(see \cite{OS} for the notion of reflection 
positivity) random fields over $\R^d$, which are invariant(i.e.,
homogeneous, stationary) with respect to the Euclidean group, has been 
looked upon as closely related to the problem of constructing (Bosonic) 
relativistic
quantum fields. In such an approach, the moments of such Euclidean random 
fields are viewed as Schwinger functions which are the analytic continuation
of the vacuum expectation value (Wightman functions) of relativistic quantum
fields to purely imaginary time.

In this section, we introduce Wightman functions associated with scalar
and vector convoluted generalized white noise Euclidean random fields.
Such kind of Euclidean random fields are solutions of certain  
stochastic partial (pseudo-)differential equations of the form $LX=F$
with $F$ a Euclidean generalized white noise and $L$ a suitable
invariant (pseudo-)differential operator. In the case where $F$ is a scalar
Gaussian white noise and $L=(-\Delta +m^2)^{\alpha}$ with $\alpha\in(0,\half]$, 
the obtained random field $X$ is a generalized free Euclidean scalar quantum 
field(see e.g. \cite{Si}). In the case that $F$ is a
quaternionic Gaussian white noise, the solution $X$ of the quaternionic
Cauchy--Riemann equation $\partial X=F$ driven by $F$ is the free
Euclidean
electromagnetic quantum field. If $F$ is non Gaussian, the corresponding 
covariant random fields can be interpreted as Euclidean quantum fields 
with some nonlinear interactions.
 
As had been investigated in \cite{AGW2} in the scalar case (see also \cite{Be}
for an axiomatic result
in the vector case), under the condition of non-Gaussian white noise, such 
Euclidean random fields in general lack the reflection positivity property. 
However,
since the Schwinger functions of such random fields can be explicitly 
calculated, we can perform the analytic continuation of the Schwinger 
functions to relativistic Wightman functions "by hand"(see \cite{AGW1,
AGW2,AIK,AW} and \cite{Go}). Using the properties of
Euclidean invariance, symmetry and real-valuedness of the Schwinger functions
on one hand, and the Osterwalder--Schrader reconstruction theorem (see 
\cite{OS}) on the other hand, we can obtain that the corresponding Wightman
functions satisfy the relativistic postulates of invariance, locality and 
hermiticity, whereas spectral property and cluster property of the 
Wightman functions can be verified directly from the derived explicit formulae.

In what follows, we only briefly review these constructions. We refer the 
reader to \cite{AGW1,AGW2,AIK,AW} and \cite{Go} for all
details.

\subsection{Scalar models}
Let $\SC (\R^d)$ be the Schwartz space of all rapidly decreasing real
valued $C^\infty$-functions on $\R^d$ and $\SC '(\R^d)$ its topological 
dual. The dual pairing is  denoted by $<\cdot,\cdot>$. Let ${\cal B}$
be the $\s$-algebra generated by all cylinder sets of $\SC '(\R^d)$. Then
$(\SC '(\R^d), {\cal B})$ is a standard measurable space.
  
By the well-known Bochner-Minlos theorem (see e.g. \cite{HKPS} or  Vol. IV
of \cite{GV}), there exists
a unique probability measure $P$ on $(\SC '(\R^d), {\cal B})$ such that its
Fourier transform satisfies

\begin{equation}
\label{3.1eqa}
\int_{\SC '(\R^d)} e^{i<\vp,\om> }dP(\om) = \exp\{\int_{\R^d}\psi(\vp(x))\}, \, \,
\vp \in \SC (\R^d)
\end{equation}  
where $\psi$ is a L\' evy-Khinchine function on $\R$ given by

\begin{equation}
\label{3.2eqa}
\psi(t) = iat-\half \s^2t^2+\int_{\R\setminus\{0\}}(e^{ist}-1-
{ist \over 1+s^2}) dM(s), \, \, t \in \R
\end{equation}
with $a, \s \in \R$ and $M$ is a non-decreasing function satisfying
$$\int_{\R\setminus\{0\}}\min(1,s^2)dM(s)<\infty.$$ 
We call $P$ \underline{a
generalized white noise measure} with L\' evy-Khinchine function $\psi$. The
associated coordinate process $F:\SC(\R^d)\times(\SC'(\R^d),{\cal B},P)
\to \R$ defined by 
$$F(\vp,\om) := <\vp,\om>, \, \, \, \vp \in \SC(\R^d), \, \om \in \SC'(\R^d)$$
is called \underline{a generalized white noise}.

Let $K: \R^d \times \R^d \to \R$ be a measurable integral kernel such
that 
$$({\cal G}\vp)(x) := \int_{\R^d} K(x,y)\vp(y)dy, \, \, \vp \in \SC(\R ^d)$$
is a linear continuous mapping from $\SC(\R^d)$ to itself. Then the 
conjugate mapping $\tilde{\cal G}: \SC'(\R^d) \to \SC'(\R ^d)$ is a 
measurable
transform from $(\SC'(\R^d),{\cal B})$ to itself. Let $P_K$ denote the 
image measure of $P$ under $\tilde{\cal G}$:
$$P_K(A) := P({\tilde{\cal G}}^{-1}A), \, \, \, A \in {\cal B}.$$
Then it is not hard to derive that

\begin{equation}
\label{3.3eqa}
\int_{\SC'(\R^d)} e^{i<\vp,\om>} dP_K(\om)
 = \exp\{\int_{\R^d}\psi(\int_{\R^d}K(x,y)\vp(y)dy)dx\}
\end{equation}
for $\vp \in \SC(\R^d)$.
The coordinate process $X:\SC(\R^d) \times (\SC'(\R^d),{\cal B},P_K)
\to \R$ given by 
$$X(\vp,\om):=<\vp,\om>, \, \, \, \vp \in \SC(\R^d), \, \om \in \SC'(\R^d)$$
is a random field. Actually, $X$ is precisely ${\tilde{\cal G}}F$ 
defined by $({\tilde{\cal G}}F)(\vp,\om):=F({\cal G}\vp,\om)$. Moreover,
$X$ is a Euclidean field if $K$ is Euclidean invariant. In this case,
we can write $K(x,y):=G(x-y)$ for some function $G$ on $\R^d$(with the
corresponding invariance property), and for the Euclidean field $X$ we have
$X=G\ast F$, i.e. $X$ is
\underline{a (Euclidean) convoluted generalized white noise}. 

Now we assume that all the moments of $M$ in (\ref{3.2eqa}) are finite,
then $\psi$ is $C^\infty$-smooth in a neighborhood 
of the origin and all the 
moments of $X$ exist. We define Schwinger functions  
of $X$ on the topological tensor product 
$\SC^{\otimes n}(\R^d)\cong\SC(\R^{dn})$ as follows

\begin{equation}
\label{3.4eqa}
S_n(\vp_1\otimes\cdots\otimes\vp_n):=\int_{\SC'(\R^d)}\prod^n_{j=1}
X(\vp_j,\om)dP_K(\om)
\end{equation}
Moreover, by using the explicit form of the right hand side of
(\ref{3.3eqa}), we can calculate the truncated Schwinger functions of
the model as follows:
 
\begin{eqnarray}
\label{3.5eqa}
S^T_n(\vp_1\otimes\cdots\otimes\vp_n)&:=&
i^{-n}{\partial^n \over \partial\l_1\cdots\partial \l_n}
\{\int_{\R^d}\psi(\sum^n_{j=1}\l_j(G\ast \vp_j)(x))dx\}\mid_{
\l_1=\cdots=\l_n=0}\nonumber  \\
&=&c_n \int_{\R^{dn}}G^{(n)}(x_1,\cdots,x_n)\prod^n_{j=1}    
\vp_j(x_j)\prod^n_{j=1}dx_j
\end{eqnarray}
for $\vp_1,\cdots,\vp_n \in \SC(\R^d)$ and $n \in \N$, where
$$c_1=a+\int_{\R\setminus\{0\}}{s^3 \over 1+s^2}dM(s)$$
$$c_2=\s^2+\int_{\R\setminus\{0\}}s^2dM(s)$$
$$c_n=\int_{\R\setminus\{0\}}s^ndM(s), \, \, n \ge 3$$
$$G^{(n)}(x_1,\cdots,x_n):=\int_{\R^d}\prod^n_{j=1}G(x-x_j)dx,
\, \, n \in \N.$$
Furthermore, taking into account that the Schwinger functions can be
expressed by partial derivatives of the right hand side of (\ref{3.3eqa})
at zero, and using a generalized chain rule,
we get the following formula
$$S_n(\vp_1\otimes\cdots\otimes\vp_n) =
\sum_{I\in\PC^n}\prod_{\{j_1,\cdots,j_k\}\in I}
S^T_k(\vp_{j_1}\otimes\cdots\otimes\vp_{j_k}), \, \,
n \in \N ,
$$
which is clearly the same relation as (\ref{2.1eqa}). 

Taking now $G$ to be the Green function $G_{\a}$, say, of the 
pseudo-differential operator $(-\D + m^2_0)^{\a}$ for the mass
$m_0>0$ and $\a \in (0,\half]$, where $\D$ is the Laplace operator
on $\R^d$, namely(e.g. in the sense of Fourier transforms of tempered
distributions)
$$
G_{\alpha}(x)=(2\pi)^{-d}\int_{{\bf R}^d}
\frac{e^{ikx}}{(|k|^2+m^2_0)^\alpha}dk,
~~~x\in{\bf R}^d,
$$
then we have Euclidean fields $X=G_{\a}\ast F$ and their Schwinger
functions and truncated Schwinger functions as defined above. To perform 
analytic continuation of $S^T_n$, we need first to represent $S^T_n$ in terms 
of a Laplace transform. In fact, we have (see \cite{AGW1,AGW2} 
and \cite{Go}) a sequence of truncated Wightman functions
$\{W^T_{n,\a}\}_{n\in\N}$, with the following Laplace transform formula

\begin{equation}
\label{3.55eqa}
S^{T}_{n}(y_1,\cdots,y_n)=(2\pi)^{-\frac{dn}{2}}
\int_{{\bf R}^{dn}}e^{-\sum^m_{l=1}k^0_ly^0_l+i\vec{k}_{l}\vec{y}_{l}}
\hat{W}^T_{n,\alpha}(k_1,\cdots,k_n)\otimes^{n}_{l=1}dk_{l}
\end{equation}
for $y^0_1<\cdots<y^0_n$, where $W^T_{1,\a}:=0$ (we take this for 
simplicity); $W_{2,\half}$ is given as $c_2$ times the two--point function 
of the relativistic free field of mass $m_0$; for $n \ge 3$ or $n=2$ and 
$\alpha \in (0,\half)$,

\begin{eqnarray}
\label{3.6eqa}
\hat{W}^T_{n,\a}(k_1,\cdots,k_n) & := &
c_n2^{n-1}(2\pi)^d\left\{\sum^n_{j=1}\prod^{j-1}_{l=1}
\mu^-_\alpha(k_l)\mu_\alpha(k_j)
\prod^n_{l=j+1}\mu^+_\alpha(k_l)\right\}\nonumber  \\ 
&{\,}&\times\delta(\sum^n_{l=1}k_l).
\end{eqnarray}
are tempered distributions with
\begin{eqnarray*}
\mu^+_\alpha(k) &:=& (2\pi)^{-d/2}\sin\pi\alpha 1_{\{k^2>m^2_0,k^0>0\}}(k)
                   \frac{1}{(k^2-m^2_0)^\alpha}  \\
\mu^-_\alpha(k) &:=& (2\pi)^{-d/2}\sin\pi\alpha 1_{\{k^2>m^2_0,k^0<0\}}(k)
                   \frac{1}{(k^2-m^2_0)^\alpha}  \\
\mu_\alpha(k) &:=& (2\pi)^{-d/2}\left(\cos\pi\alpha 1_{\{k^2>m^2_0\}}(k)+
                   1_{\{k^2<m^2_0\}}(k)\right)\frac{1}{|k^2-m^2_0|^\alpha}
\end{eqnarray*}  
where $k:=(k^0,\vec k) \in \R \times \R^{d-1}$.

By the general property of Laplace transform, $S^T_n$ can be
analytically continued from the purely Euclidean imaginary time to
the permuted extended backward 
tube $T^n_{p.e.}$ with the boundary value $W^T_{n,\alpha}=\FC^{-1}
\hat{W}^T_{n,\alpha}$ for real (relativistic) time. 
We then have the following result (see Corollary 7.11
of \cite{AGW2})

\begin{Theorem}
\label{3.1thm}
$\{W_{n,\a}\}_{n\in\N}$ defined by $\{W^T_{n,\a}\}_{n\in\N}$
via (\ref{2.1eqa}) is a sequence of Wightman functions which
satisfy Axioms I--IV, the hermiticity condition and the 
cluster property.
\end{Theorem}

\subsection{Vector models}
Euclidean vector models of quantum fields given by solutions of covariant
stochastic partial differential equations with white noise source have
been discussed in \cite{AIK}(see also references therein). We recall
here briefly the basic elements, in the case of a four dimensional
space--time, identified, in its Euclidean version, with the vector space of 
quaternions(this identification permitting to write the basic ~stochastic 
partial differential equation in a simple form). Thus, let $\H$ be the skew 
field of all quaternions with $\{{\bf 1},{\bf i},{\bf j},{\bf k}\}$
its canonical basis. Let $\SC (\R^4, \H)$ denote the Schwartz space of 
all rapidly decreasing functions from $\R^4$ to $\H$ and $\SC'(\R^4,\H)$
its topological dual. The dual pairing is denoted by $<\cdot,\cdot>$.

By the known Bochner--Minlos theorem (Vol. IV of \cite{GV}), there
exists a unique probability measure $P$ on the standard measurable space 
$(\SC'(\R^4),\H),\BC)$, where $\BC$ is the $\sigma$-algebra generated by all 
cylinder sets of $\SC'(\R^4,\H)$, with the following Fourier ~transform

$$
\int_{\SC'(\R^4,\H)}e^{i<\vp,\om>}dP(\om) = \exp\{\int_{\R^4}\psi(\vp(x))dx\},
\, \, \,  \vp \in \SC(\R^4,\H)
$$
where $\psi$ is a L\' evy-Khinchine function on $\H$ given by
\begin{eqnarray*}
\psi(x) & = & i\b x^0-\half\s_0{x^0 }^2-\half\s\vert\vec x\vert ^2 \\
 &  & -\int_{\H\setminus\{0\}}\left(1+i<x,y>_E1_{(0,1)}(\vert y\vert)-
e^{i<x,y>_E}\right)\nu(dy) 
\end{eqnarray*}  
with the condition that $\psi(x)=O(\vert x \vert ^{{4\over3}+\e})$ as 
$x \to 0$, where $x:=x^0{\bf 1}-x^1{\bf i}-x^2{\bf j}-x^3{\bf k}, \, 
(x^0,x^1,x^2,x^3)\in\R^4, \ \vec x := x-{\bf 1}x^0,
\b \in \R, \s_0,\s \in (0,\infty)$, $|x|$ denotes the Euclidean norm of
$x\in\H$ and $\nu$ is a L\' evy measure on $\H$ supported by the centre
of $\H\setminus\{0\}$(see \cite{AIK}).

In the same way as in Subsection 3.1, we can define the associated coordinate
process $F:\SC(\R^4,\H)\times(\SC'(\R^4,\H),\BC,P)\to\R$ by
$$
F(\vp,\om):=<\vp,\om> \, , \, \, \vp\in\SC(\R^4,\H), \,
\om\in\SC'(\R^4,\H).
$$
We call $F$ a $\H$-valued generalized white noise.

The covariant vector random fields were constructed in \cite{AIK} as 
solutions of the inhomogeneous quaternionic Cauchy-Riemann equation
$\partial X = F$ over $\H$, where $\partial$ is the quaternionic Cauchy-
Riemann operator defined by
$$\partial := {\bf 1}{\partial\over\partial x^0} - {\bf i}{\partial\over\partial x^1}
 - {\bf j}{\partial\over\partial x^2} - {\bf k}{\partial\over\partial x^3}.
$$

The conjugate operator $\bar{\partial}$ of $\partial$ is given by
$$\overline{\partial} := {\bf 1}{\partial\over\partial x^0} 
 + {\bf i}{\partial\over\partial x^1}
 + {\bf j}{\partial\over\partial x^2} + {\bf k}{\partial\over\partial x^3}
$$
and the Laplace operator is defined by $\D_{\H} := \partial\bar{\partial} =
\bar{\partial}\partial$. The Green function for $-\D_{\H}$ is given explicitly by 
$$g(x) = {1 \over 4\pi^2\vert x \vert^2}, \, \, \,  x \in
\H\setminus\{0\}.$$
Then the equation $\partial X=F$ is solved by the convolution $X = (-
\bar{\partial}g)\ast F$ which is the coordinate process associated to
the probability measure $P_X$ on $(\SC'(\R^4,\H),\BC)$ determined by 
the following Fourier transform  
$$\int_{\SC'(\R^4,\H)}e^{i<\vp,\om>}dP_X(\om) =
 \exp\{\int_{\R^4}\psi((g\ast\partial\vp)(x))dx\}, \, \, 
 \vp \in \SC(\R^4,\H).$$

Similarly to the scalar case in subsection 3.1, under the assumption
that $\nu$ has moments of all orders large than one, $\psi$ is 
$C^{\infty}$-smooth in a ~neighborhood of $0 \in \H$. The Schwinger
functions $S_n, n\in\N$ and the truncated Schwinger functions 
$S^T_n, n\in\N$ of $X$ can be constructed explicitly as follows
$$
S_n(\vp_1\otimes\cdots\otimes\vp_n):=\int_{\SC'(\R^4,\H)}
\prod^n_{j=1}X(\vp_j,\om)dP_X(\om) \, ,  n\in\N
$$
and 
\begin{eqnarray*}
S^T_n (\vp_1\otimes\cdots\otimes \vp_n)&:=& 
i^{-n}\frac{\partial^n}{\partial\lambda_1\cdots\partial\lambda_n}
\{\int_{\R^4}\psi(\sum^n_{j=1}\lambda_j(g\ast\partial \vp_j)(x))dx\}|_{
\lambda_1=\cdots=\lambda_n=0}  \\
&=&\left\{ \begin{array}{ll}
  constant,  &  n=1  \\
  <c_0 \, div \, \vp_1\otimes \, div \, \vp_2
                    +cD^E(\vp_1\otimes \vp_2),g^{(2)}>,  &  n=2  \\
  <\EC^n \vp_1\otimes\cdots\otimes \vp_n,g^{(n)}>,  &  n\ge 3         
\end{array} \right.
\end{eqnarray*}  
for $\vp_1,\cdots,\vp_n\in {\cal S}({\R^4,\H})$, where 
\begin{equation} 
\label{3.7eqa}
g^{(n)}(\underline{y}) =\left\{ 
\begin{array}{l}
-{1\over8\pi}\ln\vert y_1-y_2\vert, n=2;  \\ 
\int_{\R^4}\prod^n_{j=1}g(x-y_j)dx,  n\ge3.
\end{array} 
\right.
\end{equation}
for $\underline{y}=(y_1,\cdots,y_n)\in(\R^4)^n_{\ne}:=\{
\underline{y}\in(\R^4)^n:y_j\ne y_l \, \, 
if \, \, j\ne l\}$,

$$\EC^n := \sum^n_{l=0,l : even}c^n_l {\EC}^n_l,$$

\[
\left\{ \begin{array}{l}
c_0:=\s_0+\int_{\H\setminus\{0\}}{x^0}^2\nu(dx),  \\
c:=\s+{1\over3}\int_{\H\setminus\{0\}}\vert \vec x \vert^2\nu(dx),  \\
c^n_l:= \left(\begin{array}{c} n \\ l \end{array} \right) 
         {1\over l+1}\int_{\H\setminus\{0\}}{x^0}^{n-l}\vert \vec x 
        \vert^l\nu(dx), n\ge 3, 0\le l \le n, 
\end{array} \right.
\]

$${\EC}^n_l:=Sym(\underbrace{div\otimes\cdots\otimes div}_{n-l}\otimes
\underbrace{D^E\otimes\cdots\otimes D^E}_{l\over2}),
$$
and $D^E:\SC(\R^4\times\R^4,\H\times\H)\to\SC(\R^4\times\R^4,\R)$ is a
linear partial differential operator on $\R^4\times\R^4$ which is of 
first order with respect to each variable $x_1,x_2\in\R^4$.

The analytic continuation of $\{S^T_n\}_{n\in\N}$ from the imaginary
Euclidean time to the real relativistic time performed in \cite{AIK}
(to which we refer for details) yields a sequence of truncated Wightman 
functions $\{W^T_n\}_{n\in\N}$. In fact, each $g^{(n)}$ defined by 
(\ref{3.7eqa}) has a holomorphic extension $G^{(n)}$
defined on the permuted extended backward tube $T^n_{p.e.}$.
For $n\ge3$, it is defined as follows:
$$G^{(n)}(\underline{z}) := <e(\underline{z},\cdot),M^n_0+
\sum^{n-1}_{j=1}({\partial\over\partial k^0_{j+1}}-
{\partial\over\partial k^0_j})M^n_j+M^n_n>, \, \, 
\underline{z} \in T^n_{p.e.}$$
where 
$$e(\underline{z},\underline{k}):=(2\pi)^{-2n}\exp\{i\sum^n_{j=1}
<z_j,k_j>_E\}$$
and $\{M^n_j:0\le j\le n\}$ are measures defined on the space 
$\R^{4n}$ (see \cite{AIK}). This can be verified by writing $g^{(n)}$ 
as the Laplace transform (c.f. equation (\ref{3.55eqa}))
of the following tempered distribution
$$
M^n_0+\sum^{n-1}_{j=1}({\partial\over\partial k^0_{j+1}}-
{\partial\over\partial k^0_j})M^n_j+M^n_n \, .
$$
 
In what follows,
we will give a representation of $\{M^n_j:0\le j\le n\}$, which is 
different from the one given in \cite{AIK}, for 
later use in Section 4, which can be derived from the argument in 
Subsection 7.4 of \cite{AGW2} in the case $m_0=0$:

\begin{eqnarray}
\label{3.7beqa}
M^n_0 & = & (2\pi)^{3-n}{1\over 2|\vec k_1|(k^0_1-\vert\vec{k}_1\vert)}
            \prod^n_{l=2}\d^+_0(k_l)\d(\sum^n_{l=1}k_l); \nonumber \\
M^n_n & = & (2\pi)^{3-n}{1\over 2|\vec k_n|(k^0_n+\vert\vec{k}_n\vert)}
            \prod^{n-1}_{l=1}\d^-_0(k_l)\d(\sum^n_{l=1}k_l); \nonumber  \\
M^n_j (\vp) & = & \nonumber \\
&&\hspace{-.6in}(2\pi)^{3-n}\int^1_0 <\prod^{j-1}_{l=1}\d^-_0(k_l)
             {\d(k^0_j-\tilde{k^0_j}(s))\over
             4\vert\vec{k_j}\vert\vert\vec{k_{j+1}}\vert} 
             \prod^n_{l=j+2}\d^+_0(k_l)
             \d(\sum^n_{l=1}k_l),\vp>ds, \nonumber \\ 
\end{eqnarray}
for $1\le j\le n-1,\vp\in\SC(\R^{4n})$, where 
\begin{eqnarray*}
\d^+_0(k_l) & = & 1_{\{ k^0_l>0\}}(k^0_l)\d(k^2_l), \\
\d^-_0(k_l) & = & 1_{\{ k^0_l<0\}}(k^0_l)\d(k^2_l), \\
\tilde k^0_j(s) & = & \tilde k^0_j (k^0_1 , \cdots , k^0_{j-1} , \vec k_j ,
  \vec k_{j+1} , k^0_{j+1}, \cdots, k^0_n,s)\\
& := & - \{ (
  -\sum^{j-1}_{l=1} k^0_l + \om_j ) s + ( \om_{j+1}
  + \sum^n_{l=j+2} k_l^0 ) (1-s) + \sum^{j-1}_{l=1} k^0_l
  \} \ .
\end{eqnarray*}

For $n\ge3$, let $G_n$ denote the boundary value of $G^{(n)}$ 
(under the limit of the purely real time) in the backward 
tube $T^n$. For the case that $n=2$, $G_2$ can be calculated by using a
different method. Since here we do not need an explicit formula for
$W^T_2$, we refer to \cite{AGW2} for this calculation.
The corresponding truncated Wightman functions $\{W^T_n\}_{n\in\N}$
over Minkowski space $M_4$ are then given as follows

\begin{equation}
\label{3.8eqa}
W^T_n(\vp_1\otimes\cdots\otimes \vp_n) 
\left\{ \begin{array}{ll}
  \equiv  constant,  &  n=1  \\
  :=  <c_0 \, div \, \vp_1\otimes \, div \, \vp_2
                    +cD^M(\vp_1\otimes \vp_2),G_2>,  &  n=2  \\
  :=  <\LC^n \vp_1\otimes\cdots\otimes \vp_n,G_n>,  &  n\ge 3         
\end{array} \right.
\end{equation}   
for $\vp_1,\cdots,\vp_n\in\SC(\R^4,\H)$, where 
$D^M$ is a linear partial differential operator on $\R^4\times\R^4$ 
which is obtained as an analytic continuation of $D^E$ and hence it is
of first order with respect to each variable $x_1,x_2\in\R^4$, 
$$\LC^n := \sum^n_{l=0,l : even}c^n_l {\LC}^n_l,$$  
with 
$${\LC}^n_l:=Sym(\underbrace{div\otimes\cdots\otimes div}_{n-l}\otimes
\underbrace{D^M\otimes\cdots\otimes D^M}_{l\over2}) \, .
$$
We notice that ${\LC}^n_l$ is also a linear partial differential operator on 
$\underbrace{\R^4\times\cdots\times\R^4}_{n}$ which is of first
order with respect to every variable $x_1,\cdots,x_n\in\R^4$.
 
We then have the following result from Theorem 4.21 of \cite{AIK}
(cf. Theorem 4.5 and Corollary 4.7 of \cite{AGW2} for the cluster
property):
\begin{Theorem}
\label{3.2thm}
$\{W_n\}_{n\in\N}$, as defined by 
$\{W^T_n\}_{n\in\N}$ via (\ref{2.1eqa}), is a sequence of Wightman
functions which satisfy Axioms I--IV, the hermiticity condition
and the cluster property.
\end{Theorem}

\section{Verification of the Hilbert space structure condition for the
models}

In this section we prove that the truncated Wightman functions of
the scalar models as well as the vector models in Section 3 fulfill the
requirements of Corollary \ref{2.2cor}, which further implies that
the Wightman functions of both the scalar and vector models in Section 3
satisfy Axiom V. Thus, we prove the following result:

\begin{Theorem}
\label{4.1thm}
The Wightman functions obtained in Section 3 for the scalar and the
vector models(over the $d$-dimensional resp. 4-dimensional Minkowski 
space--time) fulfill the modified Wightman axioms I--V (of Morchio and
Strocchi).

In particular, for each such model there is a Hilbert space 
$(\HC,(\cdot,\cdot))$, a
continuous and self adjoint metric operator $T$ on $\HC $ fulfilling
$T^2=1$ and local $T$-symmetric field operators $\phi (f) $ defined on a common 
dense domain $\DC \subset \HC$ for $f\in \SC(\R^d,\C)$, 
$\SC (\R^4,\C^{4})$ respectively, such that equation
(\ref{XX}) holds. Furthermore, we have a $T$-unitary representation $\UC$
of (the proper orthochronous Poincar\' e group over $\R^d$ resp. $\R^4$) 
$\PC^\uparrow _+$ on the dense domain $\DC \subset \HC$, where the
transformation law of the fields $\phi$ under $\UC$ is given by (\ref{YY})
and $\UC$ fulfills the spectral condition as given in the equation (\ref{ZZ}).
\end{Theorem}

The second part of theorem \ref{4.1thm} by the results of Section 1 immediately
follows from the Axioms I-V. 

Although there is a lot of similarity in the methods applied
to the scalar and the vector model, the origin of the 
technical difficulties in the proof of Axiom V in both cases is quite
different: 

In the scalar case the K\"allen--Lehmann representation 
of the Green functions $G_\a$ by infinite measures \cite{AGW2} leads to
singularities of the Fourier transformed Wightman distributions near the
mass shell of the lowest mass. These singularities for $0<\a\leq\half$ 
turn out to be locally integrable independently of the dimension 
$d$ of the underlying space-time.

In the vector case we restricted ourselves to a special Green function,
such that the above mentioned singularities do not arise. But in this case 
we have to overcome the problems caused by the fact that the fields have
mass zero, leading to singularities at the bottom of the light cone. 
These singularities are however locally integrable, since we have specialized 
to the sufficiently large (physical) space-time dimension 4.

\subsection{Proof of Theorem 4.1 for the scalar models}
By the argument given in Section 2, it suffices to check equation
(\ref{2.5eqa}) for $n\geq 3$, $K=0$ and $N=2d$. Using the explicit
formulae of $\hat W_{n,\a}^T$ for $\a\in (0,\half ]$ with $m_0>0$,
we get that for $\vp \in \SC (\R^{dn},\C)$
\begin{eqnarray*}
|\hat W _{n,\a}^T(\vp)| &\leq & nc_n2^{n-1}(2\pi)^{d-{dn\over 2}}
\int_{\R^{d(n-1)}}\prod_{l=2}^n  {|k_l^2-m_0^2|^{-\a}\over
(1+|k_l|^2)^{d}} \\
&\times & |(\sum_{l=2}^nk_l)^2-m_0^2|^{-\a} \bigotimes_{l=2}^ndk_l \ \|
\vp \| _{0,2d}\ .
\end{eqnarray*}
It remains to show that the integral on the RHS is finite.  Noticing that 
$(1+|k_l|^2)^{-d}\leq (1+{k_l^0}^2)^{-1}(1+|\vec k_l|^2)^{1-d}$, the above 
integral can be estimated by the following expression:
\begin{eqnarray}
\label{4.1eqa}
&& \left( \int _{\R^{d-1}} {d\vec k\over (1+|\vec k|^2)^{d-1}} \right) ^{n-1} \
\left( \sup_{\vec k\in\R^{d-1}}\int _{\R} 
{|k^2-m_0^2|^{-\a}\over (1+{k^0}^2)}dk^0 \right)^{n-3}
\nonumber \\
&\times& \sup_{\begin{array}{c}{\scriptstyle \vec k_2,\vec k_3\in \R^{d-1}}\\
{\scriptstyle k_4,\cdots ,k_n\in \R^d}\end{array}}
\int _{\R^2} {|(\sum_{l=2}^nk_l)^2-m_0^2)(k_2^2-m_0^2)(k_3^2-
m_0^2)|^{-\a}\over (1+{k_2^0}^2)(1+{k_3^0}^2)} dk_2^0 dk_3^0 \ .
\nonumber \\
\end{eqnarray}
Clearly the first factor in (\ref{4.1eqa}) is finite. It remains
to show that the remaining two factors are also finite.

Since $|k^2-m_0^2|^{-\a} = |k_l^0+\om |^{-\a} |k_l^0-\om|^{-\a}$, where
$\om = (|\vec k|^2+m_0^2)^\half$ and therefore $\om \geq m_0$, we get that
\begin{equation}
\label{4.2eqa}
|k^2-m_0^2|^{-\a}\leq |m_0(k^0+\om)|^{-\a}+|m_0(k^0-\om)|^{-\a}
\end{equation}
We set $\om_1:= (|\sum_{l=2}^n\vec k_l|^2+m_0^2)^\half$, $\om_l:=(|\vec
k_l|^2+m_0^2)^\half$ for $l=2,\cdots,n$.By (\ref{4.2eqa}) the integral 
in the second factor in equation
(\ref{4.1eqa}) can be estimated by $m_0^{-\a}$ times two integrals 
of the following kind 
\begin{eqnarray*}
\int_{\R}{|k^0\pm \om|^{-\a}\over 1+{k^0}^2} dk^0 & = & \int_{\{|k^0\pm
\om|<1\}}{|k^0\pm \om|^{-\a}\over 1+{k^0}^2} dk^0+\int_{\{|k^0\pm
\om|>1\}}{|k^0\pm \om|^{-\a}\over 1+{k^0}^2} dk^0\\
&<& {2\over 1-\a}+\int_{\R}{1\over 1+{k^0}^2}dk^0 <\infty
\end{eqnarray*}
Here the latter estimate is independent of $\vec k \in \R^{d-1}$.
Consequently the second factor in (\ref{4.1eqa}) is also finite. It
remains to deal with the third factor.

Again by (\ref{4.2eqa}) the integral in the third factor of
(\ref{4.1eqa}) can be dominated by $m_0^{-\a}$ times eight
integrals of the type
$$ \int _{\R^2} {|(\sum_{l=2}^nk_l^0)\pm\om_1)(k_2^0\pm \om_2)(k_3^0\pm
\om_3)|^{-\a}\over (1+{k_2^0}^2)(1+{k_3^0}^2)} dk_2^0 dk_3^0 \ .$$
Therefore, to prove that the third factor in
(\ref{4.1eqa}) is finite it is sufficient to show that
\begin{equation}
\label{4.3eqa}
 \sup_{a,b,c\in \R} \int_{\R^2}{|xy(x+y+c)|^{-\a}\over (1+(x+a)^2)(1+(y+b)^2)}dxdy <
\infty . 
\end{equation}
To prove (\ref{4.3eqa}), we set $t:=y+c$, then we get
\begin{equation}
\label{4.4eqa}
\int_{\R} {|x(x+t)|^{-\a}\over 1+(x+a)^2} dx = \int _{\R} {|(x'+{t\over
2})(x'-{t\over 2})|^{-\a}\over 1+(x'-{t\over 2}+a)^2} dx' \ . 
\end{equation}
For the case that $|t|>2$, the RHS of (\ref{4.4eqa}) is smaller than
$$ 2\int_0^\infty {|x'-{|t|\over 2}|^{-\a}\over 1+(x'-{|t|\over 2})^2}dx'
< 2\int _{\R} {|x''|^{-\a}\over 1+{x''}^2} dx''<\infty \ 
$$
independently of $a\in \R $ and the value of $|t|>2$. We now let $0<|t|\leq 2$. In
this case the RHS of (\ref{4.4eqa}) independently of $a\in \R$ is smaller than
$$\int_{-2}^2|(x'+{t\over 2})(x'-{t\over 2})|^{-\a}dx'+\int_{\R}{1\over
1+{x''}^2}dx''\ .$$
Here the second integral is finite. For any $\g \in (0,\half)$ the first
integral can be further estimated by 
$$ 2^{1-\g} |t|^{-\g} \int _0^2 \left|x'-{|t|\over 2} \right|^{-2\a+\g} dx'\leq
2^{1-\g} |t|^{-\g}\int_{-1}^2 |x''|^{-2\a+\g}dx''\ . $$
Since $2\a-\g < 1$, the integral on the RHS of the above inequality
is finite and thus the RHS of (\ref{4.4eqa}) is smaller than
$$ C_1+C_2|t|^{-\g}$$
for sufficiently large constants $C_1,C_2>0$, which can be ~chosen 
independently of the parameter $a\in\R$. 

We can therefore estimate the left hand side(LHS) of (\ref{4.3eqa}) by 
\begin{eqnarray*}
&& \sup _{b,c\in \R} \int _{\R} {(C_1+C_2|y-c|^{-\g})|y|^{-\a}\over
1+(y+b)^2}dy\\
&\leq& C_1 \sup_{b\in \R}\int_{\R}{|y|^{-\a}\over 1+(y+b)^2}dy 
+ C_2\sup_{b,c\in \R} \int_{\R} 
{|y+c|^{-\a-\g}+|y|^{-\a-\g} \over 1+(y+b)^2}dy  \ .
\end{eqnarray*}
Here the first integral on the RHS of the above inequality is smaller than
$${2\over 1-\a}+\int _{\R}{1\over 1+{y'}^2}dy' <\infty , $$
and this estimate is independent of $b\in \R$. The
second one is dominated by the following constant
$$ {4\over 1-\a-\g} + 2 \int_{\R} {1\over
1+{y'}^2}dy' \ , $$
which is independent of $b,c\in \R $ and is finite since $\a+\g <1$. 

Thus we have established (\ref{4.3eqa}), which was the missing step
in the proof of the truncated Hilbert space structure condition
(\ref{2.5eqa}).

\subsection{Proof of Theorem 4.1 for the vector models}

As in the preceding subsection we want to prove that the requirements
of Corollary \ref{2.2cor} are fulfilled by the Fourier transformed Wightman
functions $W_n^T$ of the vector models described in Section 3. 

To this aim, we denote the Fourier transform of the partial differential
operators $\LC^n$ by $\MC^n$. $\MC^n$ is a tensor valued
multiplication operator mapping $\SC (\R^{4n},\C^{4^n})$ to $\SC
(\R^{4n},\C)$. Since $\LC^n$ is a first order partial differential
operator in the variables $x_1,\cdots ,x_n\in \R^4$, each component
of $\MC^n$ is a polynomial of degree 1 in each
of the variables $k_1,\cdots,k_n$, ~which are conjugated to 
$x_1,\cdots,x_n$ under the Fourier transform. Thus, for $K,N\in \N_0$
there exists a constant $C_1>0$, such that
\begin{equation}
\label{4.5eqa}
\| \MC^n \vp \|_{K,N}\leq C_1\|\vp\|_{K,N+1} \ \, , \forall \vp \in \SC
(\R^{4n},\C^{4^n}).
\end{equation}
By application of the Leibniz rule and the above estimate, we get that
there exists a constant $C_2>0$, such that also the ~following inequality
holds
\begin{equation}
\label{4.6eqa}
\| ({\p \over \p k_j^0}-{\p\over\p k_{j+1}^0}) \MC^n\vp \|_{K,N} 
\leq C_2 \| \vp \|_{K+1,N+1}  
\ \, , \forall \vp \in \SC (\R^{4n},\C^{4^n}),
\end{equation}
for $j=1,\cdots,n-1$. Now let again $n\geq 3$ and $\vp \in \SC
(\R^{4n},\C^{4^n})$. From (\ref{3.8eqa} ) we get
\begin{eqnarray*}
\hat W_n^T (\vp) &=& < \hat G_n,\MC^n \vp> \\
&&\hspace{-.6in} =  <M_0^n,\MC ^n\vp>+\sum_{j=1}^{n-1}<M_j^n,({\p\over \p k_j^0}-
{\p\over \p k_{j+1}^0})\MC^n\vp>+<M_n^n,\MC ^n\vp>.
\end{eqnarray*}
Taking into account the inequalities (\ref{4.5eqa}) and (\ref{4.6eqa})
we get that
$$|\hat W_n^T (\vp)|\leq a_n \| \vp \| _{K+1,N+1} \ \, , \forall \vp \in \SC
(\R^{4n},\C^{4^n})$$
for the constant $a_n:= \max\{C_1,C_2\}\sum _{j=0}^nC_j^n$, if 
the measures $M_j^n$, $j=1,\cdots,n$ fulfill the conditions
\begin{equation}
\label{4.7eqa}
|M_j^n(\vp)|\leq C_j^n\|\vp\|_{K,N} \ \forall \vp \in \SC
(\R^{4n},\C^{4^n})
\end{equation}
for sufficiently large constants $C_j^n>0$.
Thus, if we can choose $K,N\in \N_0$ in (\ref{4.7eqa}) independently 
of $n,j$, the truncated
Hilbert space condition of Corollary \ref{2.2cor} holds.

Let $K=0, N=3$. We first prove (\ref{4.7eqa}) for $j=0$: By 
(\ref{3.7beqa}) we get that
\begin{eqnarray*}
|M_0^n(\vp)|&\leq& (2\pi)^{3-n}2^{-n} \int _{\R^{3n-3}} {|\vec k_1+\cdots 
+ \vec k_n|^{-1}\over |\vec
k_2|+\cdots +|\vec k_n|+|\vec k_2+\cdots +\vec k_n|}\\
&\times& \prod _{l=2}^n{|\vec
k_l|^{-1}\over (1+|\vec k_l|^2)^{3/2}} \bigotimes_{l=2}^nd\vec k_l \ \|
\vp \| _{0,3}
\end{eqnarray*}
We have to show that the integral on the RHS is finite. Since $|\vec
k_2|+\cdots +|\vec k_n|+|\vec k_2+\cdots +\vec k_n|\geq |\vec k_3|$ the
integral is smaller than the following expression:
\begin{eqnarray}
\label{4.8eqa}
\left( \int_{\R^3} {| \vec k|^{-1}\over (1+|\vec k|^2)^{3/2}} d\vec k
\right) ^{n-3} \left( \int_{\R^3} {| \vec k|^{-2}\over (1+|\vec k|^2)^{3/2}} d\vec k
\right) && \nonumber \\
\times \left( \sup _{\vec a \in \R^3} \int _{\R^3} {|\vec k+\vec a|^{-
1}|\vec k|^{-1} \over (1+|\vec k|^2)^{3/2}} d\vec k \right) .&&
\end{eqnarray}
Let us consider the first two factors, i.e. we let $\g =1,2$ and
calculate 
$$\int _{\R^3} {|\vec k|^{-\g}\over (1+|\vec k|^2)^{3/2}}d\vec k = 4\pi
\int _0^\infty {\l^{2-\g}d\l\over (1+\l^2)^{3/2}} < \infty,$$
since $2-\g\geq 0$ and $3-2+\g>1$. It remains to show that also the
third factor in (\ref{4.8eqa}) is finite. For the moment we fix $\vec a\in
\R^3$ and choose orthogonal coordinates, such that $\vec a =(a,0,0)$. Let $\l
:= ({k^2}^2+{k^3}^2)^\half $. Using Fubini's theorem we get that the
integral in the last factor is smaller than
$$
2\pi \int _0^\infty \int_{\R} ( (k^1+a)^2+\l^2)^{-\half}( {k^1}^2+\l^2)^{-
\half}dk^1{\l d\l \over (1+\l^2)^{3/2}}
$$
Clearly $( (k^1+a)^2+\l^2)^{-\half}$ and $( {k^1}^2+\l^2)^{-\half} \in L^2(\R
, dk^1)$ for $\l >0$. By the Cauchy Schwarz inequality we can dominate
the inner integral by $\int_{\R}({k^1}^2+\l^2)^{-1}dk^1 = \pi \l^{-1}$.
Therefore, the above expression is smaller than
$$2\pi^2\int_{\R}{d\l\over (1+\l^2)^{3/2}}<\infty $$
independently of $\vec a \in \R^3$.

The estimate $|M_n^n(\vp)|\leq C_n^n\|\vp\|_{0,3}$ for a sufficiently
large $C_n^n>0$ can be proved analogously.

Let us therefore consider the case $j=1,\cdots,n-1$. From the 
representation (\ref{3.7beqa}) we get the estimate
\begin{eqnarray*}
|M_j^n(\vp)|&\leq& (2\pi)^{3-n} 2^{-n} \int _{\R^{3n-3}}
{|\sum_{l=1,l\not =j}^n\vec k_l|^{-1}|\vec k_{j+1}|^{-1}\over (1+|\vec
k_{j+1}|^2)^{3/2}} \\
&\times & \prod_{l=1,l\not = j}^n {|\vec k_l|^{-1}\over (1+|\vec
k_l|^2)^{3/2}} \bigotimes_{l=1,l\not = j}^n d\vec k_l \ \|\vp \| _{0,3}
\end{eqnarray*}
for $\vp \in \SC (\R^{dn},\C)$. The integral can be dominated by the
expression
$$
\left( \int_{\R^3} {| \vec k|^{-1}\over (1+|\vec k|^2)^{3/2}} d\vec k
\right) ^{n-2}\left( \sup _{\vec a \in \R^3} \int _{\R^3} {|\vec k+\vec a|^{-
1}|\vec k|^{-1} \over (1+|\vec k|^2)^{3/2}} d\vec k \right),
$$
which is finite by the above calculations.

This completes the proof of the truncated Hilbert space structure
condition on the truncated Wightman functions of the vector models.

\

{\bf Acknowledgements} Stimulating discussions of the first named author
with R. Gielerak, K. Iwata and T. Kolsrud are gratefully acknowledged. The  
financial support of D.F.G. (SFB 237) is also gratefully acknowledged. We 
also would like to thank the referee for very helpful comments and 
corrections on a previous version of this paper.

\end{document}